\documentclass{PoS}
\newcommand{\be}{\begin{equation}}
\newcommand{\ee}{\end{equation}}
\newcommand{\bea}{\begin{eqnarray}}
\newcommand{\eea}{\end{eqnarray}}
\newcommand{\nn}{\nonumber}

\def\({\left(}
\def\){\right)}

\newcommand{\half}{\frac{1}{2}}

\newcommand\mpl{M_{\rm Pl}}

\def\l{\left}
\def\r{\right}

\def\gapp{\mathrel{\raise.3ex\hbox{$>$}\mkern-14mu
              \lower0.6ex\hbox{$\sim$}}}

\def\lapp{\mathrel{\raise.3ex\hbox{$<$}\mkern-14mu
              \lower0.6ex\hbox{$\sim$}}}

\title{Dark Energy and Cosmology}

\ShortTitle{Dark Energy and Cosmology}

\author{\speaker{Mark Trodden}\thanks{Invited plenary presentation delivered at ICHEP 2012.}\\
        Center for Particle Cosmology, Department of Physics and Astronomy,
University of Pennsylvania, Philadelphia, PA 19104\\
        E-mail: \email{trodden@physics.upenn.edu}}

\abstract{Understanding the space of possible theoretical explanations for the observed cosmic acceleration is a central
challenge of modern cosmology. This brief document sketches selected aspects of approaches to this problem, focusing on the
possibility that a modification to General Relativity is responsible for acceleration, as presented as a plenary talk at the ICHEP 
2012 conference.}

\FullConference{36th International Conference on High Energy Physics,\\
		July 4-11, 2012\\
		Melbourne, Australia}

\begin{document}
\section{Introduction}
The phenomenon of cosmic acceleration poses a central question to fundamental physics and represents one of the outstanding open scientific questions of our time. In this article I will briefly sketch some of the ways in which researchers have approached the question of cosmic acceleration, focusing on the possibility of modifying General Relativity for the task. I will describe the difficulties in doing this in a sensible way, and focus on those ideas that I have worked on, finally briefly discussing some recent progress stemming from extra dimensional approaches connected to massive gravity. Thus, I will begin by discussing $f(R)$ models, and then briefly discuss their generalizations. I will then turn to induced gravity models, beginning with the DGP model, and then describing some generalizations and the fascinating four-dimensional effective field theories to which they reduce -- the Galileons. Finally, I will quickly comment on the general strategy for observationally distinguishing among these possibilities.

This article summarizes a plenary talk delivered at the 36th International Conference on High Energy Physics (ICHEP-2012). Because of the time constraints of the talk, and the space constraints of this article, my referencing will be necessarily idiosyncratic, and reflect a few papers that have influenced my thinking, plus some reviews, and my own work that I referred to explicitly in the talk.

\section{The Cosmological Constant and Dynamical Dark Energy}
Current observational constraints are entirely consistent with $w=-1$ -- a cosmological constant ($\Lambda$). Although $\Lambda$ is, in some sense, the simplest theoretical solution to the Dark Energy observations, the ``cosmological constant problem'' -- why is the vacuum energy so much smaller than we expect from effective-field-theory considerations? -- remains unsolved. This has led a number of authors to consider an alternative mechanism to explain $\Lambda$ arising in string theory. 

Recent developments within string theory have revealed a ``landscape'' of possible vacua generated during the compactification of eleven dimensions down to three. The ``landscape'' has a SUSY sector with degenerate zero vacuum energy; the non-SUSY sector, however, is postulated to contain $>10^{100}$ discrete vacua with finite vacuum energy densities in the range $\{-M_{p}^{4}, M_{p}^{4}\}$. Arguments based on counting vacua suggest that, in the spectrum of this extensive distribution, the existence of phenomenologically acceptable vacua is plausible. In addition, the non-SUSY vacua are metastable and allow the possibility that a region of space will be able to sample a more extended region of the landscape than if the vacua were stable, through bubble formation.

Given the complexity of the landscape, anthropic arguments have been put forward to determine whether one vacuum is preferred over another. It is possible that further development of the statistics of the vacua distribution might help to distinguish preferred vacua and extend beyond the current vacua counting approach. However, at present, in the absence of this or a way to test this proposal, a true solution to the cosmological constant problem remains elusive.

Another possibility is that dark energy is a dynamical component of the energy budget of the universe. The simplest candidate for dynamical Dark Energy is an extremely low-mass scalar field, $\phi$, with an effective potential, $V(\phi)$.  If the field is rolling slowly, its persistent potential energy is responsible for creating the late epoch of inflation we observe today. This is, in essence, a low-scale version of cosmological inflation, with the added ingredient that it is unnecessary to find an exit to the accelerating phase.

The required ``quintessence" potentials can have scaling and tracking properties that allow the scalar field energy density to evolve alongside the other matter constituents. Collectively, quintessence potentials predict that the Dark Energy density dynamically evolve in time, in contrast to the cosmological constant. Similar to a cosmological constant, however, the scalar field is expected to have extremely small density perturbations within the causal horizon, so that they contribute little to the evolution of the clustering of matter in large-scale structure.

\section{Modifying Gravity}
General Relativity (GR) is one of the best tested theories ever developed. From laboratory tests, through solar system experiments and indirect gravitational wave measurements, to the broad understanding of the expansion of the universe, the theory has performed breathtakingly well. Nevertheless, in recent years, a number of researchers have revisited the question of whether gravity might hold surprises for us, not in the high curvature regime, in which the effective field theory approach tells us GR should break down, but in the opposite, long-distance limit. As mentioned above, a primary influence behind this interest is the observed acceleration of the universe. This effect may, of course, be due to an unnaturally small cosmological constant, or a new contribution to the mass-energy of the universe (dark energy). However, a provocative possibility is that GR itself may not provide the correct set of rules with which to understand how the known matter and radiation content affects the universe on the largest scales. It may be that curvatures and length scales in the observable universe are only now reaching values at which an infrared modification of gravity can make itself apparent by driving self-acceleration (for reviews see~\cite{Copeland:2006wr,Frieman:2008sn,Silvestri:2009hh,Caldwell:2009ix,DeFelice:2010aj}).

The metric tensor, the basic element of General Relativity, contains, in principle, more degrees of freedom than the usual spin-2 graviton.
The reason why one doesn't hear of these degrees of freedom in GR is that the Einstein-Hilbert action is a very special choice, resulting in second-order equations of motion, which constrain away the scalars and the vectors, so that they are non-propagating. However, this is not the case if one departs from the Einstein-Hilbert form for the action. When using any modified action (and the usual variational principle) one inevitably frees up some of the additional degrees of freedom. In fact, this can be a good thing, in that the dynamics of these new degrees of freedom may be precisely what one needs to drive the accelerated expansion of the universe. However, there is often a price to pay.

The problems may be of several different kinds. First, there is the possibility that along with the desired deviations from GR on cosmological scales, one may also find similar deviations on solar system scales, at which GR is rather well-tested. Second is the possibility that the newly-activated degrees of freedom may be badly behaved in one way or another; either having the wrong sign kinetic terms (ghosts), and hence being unstable, or leading to superluminal propagation, which may result in other problems.

These constraints are surprisingly restrictive when one tries to create viable modified gravity models yielding cosmic acceleration. 

\subsection{A Simple Model: $f(R)$ Gravity}
The simplest modification one could think of is to replace the Einstein-Hilbert Lagrangian density by a general function $f(R)$ of the Ricci scalar $R$~\cite{Carroll:2003wy,Capozziello:2003tk}.
\be
S=\frac{\mpl^2}{2}\int d^4 x\sqrt{-g}\, \left[R+f(R)\right] + \int d^4 x\sqrt{-g}\, {\cal L}_{\rm m}[\chi_i,g_{\mu\nu}] \ ,
\label{jordanaction}
\ee
where $\mpl\equiv (8\pi G)^{-1/2}$ is the (reduced) Planck mass and ${\cal L}_{\rm m}$ is the Lagrangian
density for the matter fields $\chi_i$. Here, the matter Lagrangian is written as ${\cal L}_{\rm m}[\chi_i,g_{\mu\nu}]$ to make explicit that in
this frame - the {\it Jordan} frame - matter falls along geodesics of the metric $g_{\mu\nu}$.

The equation of motion obtained by varying the action~(\ref{jordanaction}) is
\be
\left(1+f_R \right)R_{\mu\nu} - \frac{1}{2}g_{\mu\nu}\left(R+f\right)
+ \left(g_{\mu\nu}\nabla^2 -\nabla_\mu\nabla_\nu\right) f_R
=\frac{T_{\mu\nu}}{\mpl^2} \ ,
\label{jordaneom}
\ee
where $f_R\equiv \partial f/\partial R$. Describing the matter content as a perfect fluid, with energy-momentum tensor,
\begin{equation}
T_{\mu\nu}^m = (\rho_m + p_m)U_{\mu} U_{\nu} + p_m g_{\mu\nu}\ ,
\label{perfectfluid}
\end{equation} 
where $U^{\mu}$ is the fluid rest-frame four-velocity, $\rho_m$ is the energy density and $p_m$ is the pressure, the fluid equation of motion is then the usual continuity equation. 

When considering the background cosmological evolution of such models, the metric can be taken as the flat Robertson-Walker form, $ds^2=-dt^2+a^2(t)d{\bf x}^2$. In this case, the usual Friedmann equation of GR is modified to become
\be
3H^2 -3f_R ({\dot H}+H^2)+\frac{1}{2}f+18f_{RR}H({\ddot H}+4H{\dot H})=\frac{\rho_m}{\mpl^2}
\label{jordanfriedmann}
\ee
and the continuity equation is
\be
{\dot \rho}_m +3H(\rho_m+p_m)=0 \ .
\label{jordancontinuity}
\ee
When supplied with an equation of state parameter $w$, the above equations are sufficient to solve for the
background cosmological behavior of the space-time and its matter contents. For appropriate choices of the
function $f(R)$ it is possible to obtain late-time cosmic acceleration without the need for dark energy, 
although evading bounds from precision solar-system tests of gravity turns out to be a much trickier
matter, as we shall see.

It is convenient to perform a carefully-chosen conformal transformation on the metric, in order to render the gravitational action in the usual Einstein Hilbert form of GR. We therefore write
\be
{\tilde g}_{\mu\nu} = \Omega(x^{\alpha}) g_{\mu\nu} \ ,
\label{conftrans}
\ee
and construct the function $r(\Omega)$ that satisfies
\be
1+f_R[r(\Omega)]=\Omega \ .
\ee
Defining a rescaled scalar field by $\Omega \equiv e^{\beta\phi}$, with 
$\beta\mpl\equiv\sqrt{2/3}$, the resulting action becomes
\bea
{\tilde S}=\frac{\mpl}{2}\int d^4 x\sqrt{-{\tilde g}}\, {\tilde R} &+&\int d^4 x\sqrt{-{\tilde g}}\, 
\left[-\frac{1}{2}{\tilde g}^{\mu\nu}(\partial_{\mu}\phi)\partial_{\nu}\phi -V(\phi)\right] \nonumber \\ 
&+&
\int d^4 x\sqrt{-{\tilde g}}\, e^{-2\beta\phi} {\cal L}_{\rm m}[\chi_i,e^{-\beta\phi}{\tilde g}_{\mu\nu}]\ ,
\label{einsteinaction}
\eea
where the potential $V(\phi)$ is determined entirely by the original form~(\ref{jordanaction}) 
of the action and is given by
\be
V(\phi)=\frac{e^{-2\beta\phi}}{2}\left\{e^{\beta\phi}r[\Omega(\phi)] - f(r[\Omega(\phi)]) \right\} \ .
\label{einsteinpotential}
\ee

The equations of motion in the Einstein frame are much more familiar than those in the Jordan frame, 
although there are some crucial subtleties. In particular, note that in general, test particles of the matter content $\chi_i$ do not freely fall along geodesics of the metric ${\tilde g}_{\mu\nu}$.

The equations of motion in this frame are those obtained by varying the action with respect to the metric ${\tilde g}_{\mu\nu}$
\be
{\tilde G}_{\mu\nu} = \frac{1}{\mpl^2}\left({\tilde T}_{\mu\nu} + T^{(\phi)}_{\mu\nu}\right) \ ,
\label{einsteineom}
\ee
with respect to the scalar field $\phi$
\be
{\tilde \nabla^2}\phi = -\frac{dV}{d\phi}(\phi) \ ,
\label{scalareom}
\ee
and with respect to the matter fields $\chi_i$, described as a perfect fluid.

Once again, we specialize to consider background cosmological evolution in this frame. The 
Einstein-frame line element can be written in familiar FRW form as
\be
ds^2 =-d{\tilde t}^2+{\tilde a}^2({\tilde t})d{\bf x}^2 \ ,
\label{einsteinFRWmetric}
\ee
where $d{\tilde t}\equiv\sqrt{\Omega}\, dt$ and ${\tilde a}(t)\equiv\sqrt{\Omega} \,a(t)$. The Einstein-frame matter energy-momentum tensor is then given by
\be
{\tilde T}_{\mu\nu}^m = ({\tilde \rho}_m + {\tilde p}_m){\tilde U}_{\mu} {\tilde U}_{\nu} + 
{\tilde p}_m {\tilde g}_{\mu\nu}\ ,
\label{einsteinperfectfluid}
\ee
where ${\tilde U}_{\mu}\equiv \sqrt{\Omega} \,U_{\mu}$, ${\tilde \rho}_m\equiv \rho_m/\Omega^2$ and 
${\tilde p}_m\equiv p_m/\Omega^2$.

Now, as I mentioned in the introduction, any modification of the Einstein-Hilbert action must, of
course, be consistent with the classic solar system tests of gravity
theory, as well as numerous other astrophysical dynamical tests. 
We have chosen the coupling constant $\mu$ to be very small, but
we have also introduced a new light degree of freedom.  As shown by
Chiba~\cite{Chiba:2003ir}, the simple model above is equivalent to a Brans-Dicke
theory with $\omega=0$ in the approximation where the potential
was neglected, and would therefore be inconsistent with solar system measurements~\cite{Bertotti:2003rm}.

To construct a realistic $f(R)$ model~\cite{Hu:2007nk} requires a more complicated function, with more than one
adjustable parameter in order to fit the cosmological data~\cite{Bean:2006up} and satisfy solar system bounds through the
chameleon~\cite{Khoury:2003aq} mechanism. 

\subsection{Extensions: Higher-Order Curvature Invariants}
It is natural to consider generalizing the action of~\cite{Carroll:2003wy} to include other curvature invariants~\cite{Carroll:2004de}.
There are, of course, any number of terms that one could consider, but for simplicity, focus on
those invariants of lowest mass dimension that are also parity-conserving $P \equiv  R_{\mu\nu}\,R^{\mu\nu}$ and 
$Q \equiv  R_{\alpha\beta\gamma\delta}\,R^{\alpha\beta\gamma\delta}$.

The action then takes the form
\begin{equation}
S=\int d^4x \sqrt{-g}\,[R+f(R,P,Q)] +\int d^4 x\, \sqrt{-g}\,
{\cal L}_M \ ,
\label{genaction}
\end{equation}
where $f(R,P,Q)$ is a general function describing deviations from general relativity.

Actions of the form~(\ref{genaction}) generically admit a maximally-symmetric solution that is often unstable to another accelerating power-law attractor. 
It has been shown that solar system constraints, of the type I have described for $f(R)$ models, can be evaded by these more general 
models when, for example, the $Q$ terms are relevant on those scales. However, these theories generically contain ghosts and/or superluminally propagating modes~\cite{Chiba:2005nz,DeFelice:2006pg,Calcagni:2006ye}. I therefore will not discuss them further here.

\subsection{Induced Gravity Models}

In the Dvali-Gabadadze-Porrati (DGP) model~\cite{Dvali:2000hr}, our observed $4D$ universe is embedded in an infinite empty fifth dimension. 
Despite the fact that the extra dimension is infinite in extent, the inverse-square law is nevertheless recovered at short distances on the brane due to an intrinsic, four-dimensional Einstein-Hilbert term in the action
\be
	S_{\rm DGP} = \int_{\rm bulk} {\rm d}^5x\sqrt{-g_5}\frac{M_5^3}{2}R_5 +
\int_{\rm brane} {\rm d}^4x \sqrt{-g_4} \left(\frac{M_4^2}{2}R_4 + {\cal L}_{\rm matter}\right)\,.
\ee
The Newtonian potential on the brane scales as $1/r$ at short distances, as in $4D$ gravity, and asymptotes to $1/r^2$ at large distances, characteristic of $5D$ gravity. The cross-over scale $m_5^{-1}$ between these two behaviors is set by the bulk and brane Planck masses ($M_{5}$ and $M_{4}$ respectively) via $m_5 = \frac{M_5^3}{M_4^2}$.

In this picture, the higher-dimensional nature of gravity affects the $4D$ brane through deviations from general relativity on horizon scales, that may give rise to the observed accelerated expansion. This model faces its own challenges however. The branch of solutions that include self-acceleration suffers from ghost-like instabilities, and on the observational front DGP cosmology is statistically disfavored in comparison to $\Lambda$CDM and is significantly discordant with constraints on the curvature of the universe. 

\section{Galileons and Massive Gravity}
Careful studies of the DGP model have, however, given rise to new ideas about how to construct four-dimensional effective field theories with symmetries that may be relevant for cosmology. The decoupling limit of DGP consists of a $4$-dimensional effective theory of gravity coupled to a single scalar field $\pi$, representing the bending mode of the brane in the fifth dimension. The $\pi$ field self-interaction includes a cubic self-interaction $\sim (\partial\pi)^2\square\pi$, which has the properties that the field equations are second order, and the terms are invariant up to a total derivative under the internal galilean transformations 
\be
\pi\rightarrow \pi+c+b_\mu x^\mu \ ,
\ee
where $c,b_\mu$ are arbitrary real constants.

In \cite{Nicolis:2008in}, this was generalized, and all possible lagrangian terms for a single scalar with these two properties were classified in all dimensions.  They are called galileon terms, and there exists a single galileon lagrangian at each order in $\pi$, where ``order" refers to the number of copies of $\pi$ that appear in the term.   
For $n\geq 1$, the $(n+1)$-th order galileon lagrangian is
\be
\label{galileon2} 
{\cal L}_{n+1}=n\eta^{\mu_1\nu_1\mu_2\nu_2\cdots\mu_n\nu_n}\left( \partial_{\mu_1}\pi\partial_{\nu_1}\pi\partial_{\mu_2}\partial_{\nu_2}\pi\cdots\partial_{\mu_n}\partial_{\nu_n}\pi\right),
\ee 
where 
\be
\label{tensor} 
\eta^{\mu_1\nu_1\mu_2\nu_2\cdots\mu_n\nu_n}\equiv{1\over n!}\sum_p\left(-1\right)^{p}\eta^{\mu_1p(\nu_1)}\eta^{\mu_2p(\nu_2)}\cdots\eta^{\mu_np(\nu_n)} \ .
\ee 
The sum in~(\ref{tensor}) is over all permutations of the $\nu$ indices, with $(-1)^p$ the sign of the permutation.  The tensor~(\ref{tensor}) is anti-symmetric in the $\mu$ indices, anti-symmetric the $\nu$ indices, and symmetric under interchange of any $\mu,\nu$ pair with any other.  These lagrangians are unique up to total derivatives and overall constants.   Because of the anti-symmetry requirement on $\eta$, only the first $n$ of these galileons are non-trivial in $n$-dimensions.  In addition, the tadpole term, $\pi$, is galilean invariant, and we therefore include it as the first-order galileon.  

Thus, at the first few orders, we have 
\bea
 {\cal L}_1&=&\pi, \\ \nn
 {\cal L}_2&=&[\pi^2], \\ \nn
{\cal L}_3&=&[\pi^2][\Pi]-[\pi^3], \\ \nn
{\cal L}_4&=&\half[\pi^2][\Pi]^2-[\pi^3][\Pi]+[\pi^4]-\half[\pi^2][\Pi^2], \\ \nn
{\cal L}_5&=&{1\over 6}[\pi^2][\Pi]^3-{1\over 2}[\pi^3][\Pi]^2+[\pi^4][\Pi]-[\pi^5]+{1\over 3}[\pi^2][\Pi^3]-{1\over 2}[\pi^2][\Pi][\Pi^2]+{1\over 2}[\pi^3][\Pi^2]
\ .
\eea
We have used the notation $\Pi$ for the matrix of partials $\Pi_{\mu\nu}\equiv\partial_{\mu}\partial_\nu\pi$, and $[\Pi^n]\equiv Tr(\Pi^n)$, e.g. $[\Pi]=\square\pi$, $[\Pi^2]=\partial_\mu\partial_\nu\pi\partial^\mu\partial^\nu\pi$, and $[\pi^n]\equiv \partial\pi\cdot\Pi^{n-2}\cdot\partial\pi$, i.e. $[\pi^2]=\partial_\mu\pi\partial^\mu\pi$, $[\pi^3]=\partial_\mu\pi\partial^\mu\partial^\nu\pi\partial_\nu\pi$.  The above terms are the only ones which are non-vanishing in four dimensions.  The second is the standard kinetic term for a scalar, while the third is the DGP $\pi$-lagrangian (up to a total derivative).

The first few orders of the equations of motion are 
\bea{\cal E}_1&=&1, \\
 {\cal E}_2&=&-2[\Pi], \\
{\cal E}_3&=& -3\left([\Pi]^2-[\Pi^2]\right),  \\
{\cal E}_4&=& -2\left([\Pi]^3+2[\Pi^3]-3[\Pi][\Pi^2]\right),  \\
{\cal E}_5&=& -{5\over 6}\left([\Pi]^4-6[\Pi^4]+8[\Pi][\Pi^3]-6[\Pi]^2[\Pi^2]+3[\Pi^2]^2\right) \ .
\eea

These galileon actions can be generalized to the multi-field case, where there is a multiplet $\pi^I$ of fields~\cite{Deffayet:2009mn,Deffayet:2010zh,Padilla:2010de,Hinterbichler:2010xn}.  The action in this case can be written
\be 
\label{generaltermspre} 
{\cal L}_{n+1}= S_{I_1I_2\cdots I_{n+1}}\eta^{\mu_1\nu_1\mu_2\nu_2\cdots\mu_n\nu_n}\left(\pi^{I_{n+1}} \partial_{\mu_1}\partial_{\nu_1}\pi^{I_1}\partial_{\mu_2}\partial_{\nu_2}\pi^{I_2}\cdots\partial_{\mu_n}\partial_{\nu_n}\pi^{I_{n}}\right),
\ee
with $S_{I_1I_2\cdots I_{n+1}}$ a symmetric constant tensor.  This is invariant under under individual galilean transformations for each field, $\pi^I\rightarrow \pi^I+c^I+b^I_\mu x^\mu$, and the equations of motion are second order,
\be 
{\cal E}_{I}\equiv {\delta{\cal L}\over \delta \pi^I}=(n+1)S_{II_1I_2\cdots I_{n}}\eta^{\mu_1\nu_1\mu_2\nu_2\cdots\mu_n\nu_n}\left(\partial_{\mu_1}\partial_{\nu_1}\pi^{I_1}\partial_{\mu_2}\partial_{\nu_2}\pi^{I_2}\cdots\partial_{\mu_n}\partial_{\nu_n}\pi^{I_{n}}\right) \ .
\ee
These extended models have their own constraints~\cite{Andrews:2010km}, which I won't have time to discuss further here.

The theory containing these galilean-invariant operators is not renormalizable, i.e. it is an effective field theory with a cutoff $\Lambda$, above which some UV completion is required.  However, crucially, the ${\cal L}_n$ terms above do not get renormalized upon loop corrections, so that their classical values can be trusted quantum-mechanically

In principle, one should consider quantum effects within the effective theory, since there are other operators of the same dimension that might compete with the galileon terms.  However, fascinatingly, there can exist interesting regimes where non-linearities from the galileon terms are important, yet quantum effects are under control. This separation of scales allows for the existence of regimes in which there exist classical field configurations with non-linearities of order one, and still trust these solutions in light of quantum corrections. These non-linear, quantum-controlled regimes are where interesting models of inflation, cosmology, modified gravity, etc. employing these galileon actions should be placed.

Perhaps the most interesting aspect of Galileons is that they have recently been shown to arise in a particular limit of the recently proposed
construction of the theory of ghost free massive gravity. At the linearized level around flat space, there exists a unique mass term -- the Fierz-Pauli term -- for a graviton that avoids the existence of a ghost. However, a powerful no-go theorem due to Boulware and Deser~\cite{Boulware:1973my} seemingly shut off all possible ghost free nonlinear completions of such a theory. However, in the last two years a loophole to this argument has emerged through the work of de Rham, Gabadadze and Tolley~\cite{deRham:2010ik,deRham:2010kj}. This construction is currently allowing many researchers to consider whether massive gravity alone might provide the self-consistent infrared modification of gravity necessary to tackle the cosmological constant problem and cosmic acceleration. Galleons emerge in a limit, and so might be relevant to that task. On the other hand, Galileons may be useful as an interesting set of scalar field theories in their own right.

In this context, while various authors have considered the uses of these theories for early and late cosmology, it would be interesting to find examples of theories with the same attractive symmetry features, but which naturally live on the curved manifolds relevant to cosmology. There are obstacles to performing this in a straightforward manner~\cite{Deffayet:2009wt}. However, in very recent
papers~\cite{Goon:2011qf,Goon:2011uw} we have demonstrated how to construct such models\footnote{This construction has connections to some other recently proposed ones~\cite{Deffayet:2011gz,Burrage:2011bt}}, and even how to construct a master theory describing covariant Galileons and massive gravity~\cite{Gabadadze:2012tr}.

\section{Brief Comments on Observational Prospects}
An impressive amount of work has been done on the potential for upcoming missions to distinguish among a cosmological constant, dynamical dark energy and modifications of gravity as potential explanations for cosmic acceleration. This is not the place to review these in detail, but the general idea is that the degeneracy among these ideas in terms of describing the expansion history of the universe is typically broken at the level of cosmological structure formation.

Scalar metric perturbations in the Newtonian gauge are described by two potentials, $\Psi(\vec{x},t)$ and $\Phi(\vec{x},t)$, defined via the perturbed FRW line element
\be\label{metric}
ds^2=-\l(1+2\Psi(\vec{x},t)\r)dt^2+a^2(t)\l(1-2\Phi(\vec{x},t)\r)d\vec{x}^2 \ .
\ee
These potentials correspond, respectively, to perturbations in the time-time and space-space components of the metric tensor, representing the strength of gravity and the spatial curvature. In the $\Lambda$CDM model, the potentials are equal during the epoch of structure formation, and their time dependence is set by the same scale-independent linear growth function that describes the growth of matter density perturbations. 
This, it turns out, is a very peculiar feature of the $\Lambda$CDM model, and of models of dark energy with negligible shear and clustering. However, it no longer holds in theories of modified gravity or models of coupled dark energy, where one can have scale-dependent growth patterns. In these theories, the two Newtonian potentials generically differ, and their dependence on matter perturbations can be different. 
As a consequence, modifications of gravity generally introduce a time- and scale-dependent slip between the Newtonian potentials, as well as a time- and scale-dependent effective Newton's constant describing the clustering of dark matter.  These modifications are expected to leave distinct imprints on the large scale structure of the universe, which may help to break the degeneracy that characterizes models of cosmic acceleration at the background level. A number of observables are useful for this process, including galaxy counts, which probe the distribution and growth of matter inhomogeneities, and peculiar velocity surveys, which typically use redshift-independent distance indicators to separate the Hubble flow from the local flow. In contrast to these two observables, which respond to one of the metric potentials,  namely $\Psi$, weak lensing of distant light sources by intervening structure is determined by spatial gradients of the sum ($\Phi+\Psi$). Hence, measurements of the weak lensing shear distribution over multiple redshift bins can provide an estimate of the space and time variation of the sum of the two potentials. 

By combining multiple redshift information from these three methods, and their cross-correlations, one can constrain the differences between the metric potentials and the space-time variation of the effective Newton constant. Ideally, experimentalists would measure all possible cross-correlations, between all possible pairs of observables, in order to maximize the amount of information available to us. In practice, however, it can be difficult to obtain these cross-correlations, although this will be addressed with near and distant future tomographic large scale structure surveys (such as DES, LSST and PAN-STARRS). 

To summarize, the predictions for the growth of structure in models of modified gravity and coupled dark energy are significantly different than those in the $\Lambda$CDM or uncoupled quintessence models. Therefore, in principle, large scale structure and microwave background data may allow a distinction between these two approaches to the phenomenon of cosmic acceleration.


\section{Conclusions} 

Cosmic acceleration is clearly telling us something fundamental about gravity and its coupling to matter. The cosmological constant is a great fit to all data, but a complete understanding of the required magnitude still eludes us. Dynamical dark energy is another possibility, as is the chance that general relativity may become modified on the largest scales. I have found this possibility particularly intriguing in recent years, and in this talk I  outlined a number of modern approaches to this problem, focusing on those that I have been involved with in one way or another. In the final part of the talk I discussed how higher dimensional constructions, such as the DGP model, have yielded an interesting set of four dimensional effective field theories - the {\it galileons} - encapsulating the effects of modifying gravity. It is too early to know if any of these approaches will turn out to be successful, but interesting progress is being made in understanding whether they pass increasing theoretical and observational constraints, and current and future missions have the potential to distinguish between these approaches.

\acknowledgments
I would like to thank the organizers of ICHEP 2012 for inviting me to deliver this presentation, and for a terrific meeting in Melbourne. I would also like to thank my collaborators, from whose joint work with me I have borrowed liberally in putting together this summary. This work is supported in part by NASA ATP grant NNX11AI95G, the US Department of Energy and by the Fay R. and Eugene L. Langberg chair.

\end{document}